\documentclass[12pt,preprint]{aastex}

\shorttitle{Jet triggered Type Ia supernovae in radio-galaxies?}

\shortauthors{Alessandro Capetti}

\begin{document}

\title{Jet triggered Type Ia supernovae in radio-galaxies?
\thanks{Based on observations with the NASA/ESA Hubble Space Telescope, 
obtained at the
Space Telescope Science Institute, which is operated by AURA, Inc.,
under NASA contract NAS 5-26555 and by STScI grant GO-3594.01-91A}}

\author{Alessandro  Capetti}
\affil{Istituto  Nazionale  di   Astrofisica  (INAF),\\
Osservatorio Astronomico di Torino, Strada Osservatorio 20, 10025 Pino
Torinese, Torino, Italy}

\begin{abstract}
We  report the  serendipitous discovery  of  a supernova  (SN) in  the
nearby  radio-galaxy  3C  78.
Observations obtained  with  the  STIS
spectrograph on board  the Hubble Space Telescope show,  at a distance
of  0.54 arcsec  (300 pc)  from the  galaxy nucleus,  a  second bright
source,  not   present  in  previous  images.   As   this  source  was
fortuitously  covered  by  the  spectrograph  slit  its  spectrum  was
obtained and it is characteristic of a Type Ia SN.  This SN is closely
aligned with the  radio-jet of 3C 78.  Analysis  of historical records
shows that such a close association between jet and supernova occurred
in 6 of  the 14 reported SNe in  radio-galaxies.  The probability that
this results  from a random distribution  of SN in the  host galaxy is
less than $\sim$ 0.05\%. 
We then argue that jets  might trigger supernova explosions.

\end{abstract}


\section{Introduction}
\label{intro}
Type Ia supernovae are in  many aspects still enigmatic objects. They
are  thought to be  the result  of the  thermonuclear explosion  of an
evolved degenerated star,  very likely a white dwarf  (WD) composed of
carbon  and oxygen, accreting  material from  a companion  star (Livio
2000).  However, the precise nature of the progenitor systems of SN Ia
and of  the processes of  mass accretion remain largely  unknown after
decades  of research.  In the  most accepted  scenario  (the so-called
single degenerate model), the transfer of mass from the companion star
causes  the WD  to  exceed  the Chandrasekhar  limit,  leading to  the
explosion.  In elliptical galaxies,  the age of the stellar population
restricts the  identification of  the companion star  with a  low mass
star, evolved  to the red-giant  phase whose envelope fills  its Roche
lobe  and  sustains   the  mass  transfer  onto  the   WD  (Nomoto  et
al.  2000).  However, alternative  models  are  also  viable, such  as
explosions of  sub-Chandrasekhar systems  or the possibility  that the
companion star is also degenerate (the double degenerate model) with a
combined  mass of  the system  exceeding the  critical mass:  when the
period of  the binary  system is sufficiently  small the two  WD stars
will merge due to orbital energy loss by gravitational radiation.

The association between SNe and jets of radio-galaxies 
discussed in this Letter 
opens an unexpected link between these phenomena that can be used
to shed new light on the physics and evolution of these systems.

\section{Observations}
\label{observations}
Observations of  the  nearby radio-galaxy  3C  78 (NGC 1218), at  a
redshift of z = 0.0287, were obtained on September 6th, 2000 using the
STIS  spectrograph on  board  the Hubble  Space  Telescope (HST).  The
acquisition image, taken  to accurately locate the centre of
the galaxy within the slit, shows the presence, at a distance
of 0.54  arcsec (300 pc adopting a Hubble  constant of H$_0$=75 km
s$^{-1}$ Mpc$^{-1}$) from  the nucleus, of a  second bright source
(see Fig.  1) that was  not present in  previous images of 3C  78. The
slit (0$\farcs$2  of width, 50$\arcsec$ of length) was not
constrained  to a  pre-selected orientation but,  fortuitously, it
included  also this  second source.  Spectra were  obtained  using two
low-resolution  gratings  (G430L and  G750L)  covering the  wavelength
range 3000 - 11000 \AA\ at a spectral  resolution varying between 500
and 1000 with an exposure time of 6 minutes for each grating. The data
reduction was  performed using  the standard calibration  pipeline for
STIS  data (Leitherer 2001). 

\section{Supernova classification and dating}
\label{dating}
The spectrum of the off-nuclear  source, is presented in Fig. 2.
It lacks of
any  detectable H  line while  it is  dominated by  blend of  Fe  and Co
lines. This  spectrum is typical of  a SN  Type Ia in  its nebular
phase (e.g. Filippenko 1997).

The homogeneity  of the spectral  and photometric evolution of  SNe Ia
can be used to establish the epoch of this supernova explosion, which,
as there was no report of  this event, is unknown. Note, first of all,
that the supernova was not present  in HST images of 3C 78 obtained on
March 15th, 2000.  The comparison of  the SN spectrum with those of SN
1994D  (Patat  et al.  1996),  a prototypical  SN  Type  Ia, taken  at
different  epochs and  in particular  the strength  of the  Fe  and Co
lines, suggests an age of approximately 20-40 days from the brightness
peak.

Similarly,  the characteristic  light  curves  of SN  Ia  allow us  to
measure the time elapsed from the maximum brightness. We estimated the
magnitude  of the  SN convolving  its spectrum  with  the transmission
curve of the standard B filter and estimating the slit losses from the
acquisition image, which yields B  = 18.4 $\pm$ 0.2. With its redshift
of 0.0287 the  distance modulus of 3C 78 is 35.3  from which we derive
an  absolute magnitude  for the  SN  of B=-16.9.   This luminosity  is
typical of  the end  of the peak  phase just  before the onset  of the
exponential decay  of the SNe Type  Ia light curves,  which is reached
after  about  20-30  days  from  the brightness  peak  (Leibundgut  et
al. 1991), in good agreement with the spectral dating.

\section{Supernovae in radio-galaxies} 
\label{sninrg}
The SN  discovered in 3C  78 is closely  aligned with the jet  of this
radio-galaxy: the jet is oriented at 51$^{\circ}$ from North (Unger et
al. 1984)  while the SN  is located along position  angle 42$^{\circ}$
and it lies along the outer edge of the jet.

Analysis of historical records of  SNe (Barbon et al. 1999) produces a
list  of  13  events  occurred  in  radio-galaxies  (defined  as
elliptical  or   S0  galaxies   belonging  to  catalogues   of  bright
radio-sources, i.e. 3C, 4C, B2 or PKS) prior to the one reported here.
Eight of them were classified as Type Ia, three generically as Type
I and  only two have  no classification.  Nonetheless, as  all galaxies
under examination  here are elliptical  or S0, we can  safely conclude
that they  were all  SN Ia  as this is  the only  SN type  observed in
galaxies of these  Hubble types (van den Bergh  \& Tammann 1991).  In
Table 1 we give the optical and radio identifications, the SN name and
classification  and  their  relative  position  with  respect  to  the
galaxy's nucleus.

In  8  cases  the  location  of  the SN  does  not  show  any  special
relationship  with the radio  structure. Conversely,  SN 1968A  in NGC
1275, two  SNe found in NGC  4374 (SN1957B and SN1991bg),  SN 1986G in
NGC  5128  and  SN2001ic  in  NGC  7503, are  all  located  along  the
radio-jets, with  an alignment within less than  10$^{\circ}$ from the
jet axis similarly to what is observed in 3C 78. In one case, SN
1981G in NGC 5127, although aligned  with jet position angle the SN is
not associated with it being located at a distance of 14\arcsec \ from
the nucleus, larger than the radio jet's length that is $\sim$ 10\arcsec; this
was clearly not considered as a SN associated with the jet.
In Fig. \ref{offset}
we show  the offset  between the position  angle of all  14 supernovae
with respect  to the  nearest of  the two jets  that reveals  a strong
concentration of SN  events close to the radio axis.   

The  probability of  a  fortuitous superposition  between  jet and  SN
depends  on  the  fraction  of  galaxy's  mass  covered  by  the  jets
projection.   This  estimate is  straightforward for  a spherical
galaxy, or more generally for a galaxy with circular isophotes, and if
no extinction  is present:  adopting a  
value  of 20$^{\circ}$  for the
jets opening angle, typical of the low luminosity radio-galaxies 
of Table 1 (see Parma et al. 1987) 
this association  is expected in 1/9 events.  While
indeed no  significant dust  absorption is in  general present  in the
hosts  of radio-galaxies  and  although  they are  in  general of  low
ellipticity, a small, but systematic, underestimate might arise if the
jets were aligned with major  axis of the galaxy.  However, studies of
the  relative position  angle of  optical and  radio structure  in low
redshift radio-galaxies  show that their  offset is consistent  with a
random distribution  and that no  preferential radio/optical alignment
is present (Baum \& Heckman 1989, Birkinshaw \& Davies 1985).

We can then  estimate that the odds  of finding 6 (or more)  out of 14
SNe spatially associated with the jets is 0.25\%.  This probability is
further  reduced  noting that,  in  many  cases,  the radio jets  length  is
significantly  smaller  than  the  galaxy's size.   For  example,  the
radio-jet in 3C 78 is only 2$\arcsec$ long; analysis of the HST images
indicates that within this radius  is contained only about 20\% of the
galaxy's light,  reducing the  overall probability by  a factor  of 5,
down to 0.05\%.
Thus  the observed  spatial association  is  very unlikely  to be  the
result of a  by chance superposition and suggests  a causal connection
between the SN and the presence of the jet.

\section{Conclusions}
\label{discussion}

The  spatial association  of SNe  with the  jets originating  from the
active nucleus of radio-galaxies  suggests that jets play an important
role  in  perturbing the  SN  progenitor  system  and its  environment,
leading to  an increased probability that  the SN event  occurs in the
jet's vicinity.

At this stage it is difficult to isolate the precise 
mechanism at work, but our results seem to indicate 
that the presence of a relativistic
jet might lead to a  substantial increase in the mass flow from 
either the interstellar medium (see Livio, Riess \& Sparks 2002) 
or the donor star onto the white dwarf.
This,  in turn,  would increase the  probability that  the SN
explosion occurs in the vicinity of the jet.
Clearly, if confirmed, 
the association between SNe and jets can  be used
to shed new light on the physics and evolution of the
SN progenitor systems.

From an observational  point of view, what is needed  at this stage is
an aggressive  campaign of  optical monitoring of  radio-galaxies with
the aim of providing better  statistics for the connection between jet
and SN. The SN rate in elliptical and S0 galaxies, expressed in number
of SNe per century per 10$^{10}$  solar luminosities in the B band, is
estimated at 0.67 (Evans et al.  1989).  Adopting a typical luminosity
of  L$_{\rm  B} \sim  10^{11}  $L$_{\sun}$  for  the giant  elliptical
galaxies hosting  the radio-source, a  rate of approximately 7  SN per
year is  expected in a sample  of 100 nearby objects,  formed e.g.  by
all PKS, B2 and 3C radio  sources with z $<$  0.05. This will
lead  to  an  increase of  a  factor  2  of  the  known SN  events  in
radio-galaxies in two years.
With this improved statistics it will be possible first of all to test
if the association between jet and SN is confirmed and furthermore
to establish if radio-galaxies show indeed an increased SN rate with respect
to non active galaxies.

\acknowledgments
I'd like to thank the anonymous referee for suggesting several points of 
clarification. 
This  research has made  use of  the NASA/IPAC  Extragalactic Database
(NED) which  is operated by the Jet  Propulsion Laboratory, California
Institute of Technology, under  contract with the National Aeronautics
and Space Administration.

\clearpage

\begin{figure}
\plotone{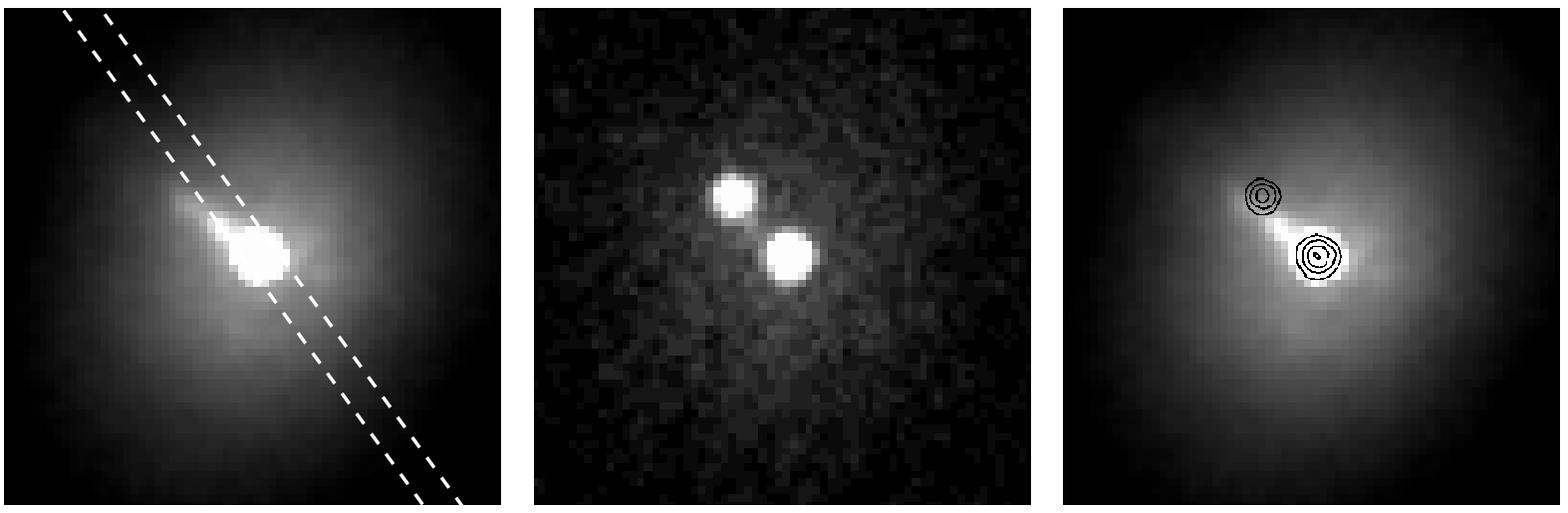}
\caption{The left panel shows the central 
portion (3.3 x  3.3 arcsec, 1.8 x 1.8  kpc) of a 280 s  Wide Field and
Planetary Camera 2 image of 3C 78 obtained on August 17th 1995 through
the F702W (R band) filter.  Clearly visible are the bright nucleus and
the optical  jet, while the diffuse  emission is the  starlight of the
host galaxy. The two parallel  dashed lines mark the slit position. In
the middle panel  we present the shorter, 10 s  of exposure time, STIS
acquisition image where, besides  the nucleus, a second bright source,
the  supernova, is present.  The right  panel shows  the superposition
(contours) of the SN location with respect to the nucleus and the jet.
\label{fig1}}
\end{figure}

\clearpage

\begin{figure}
\plotone{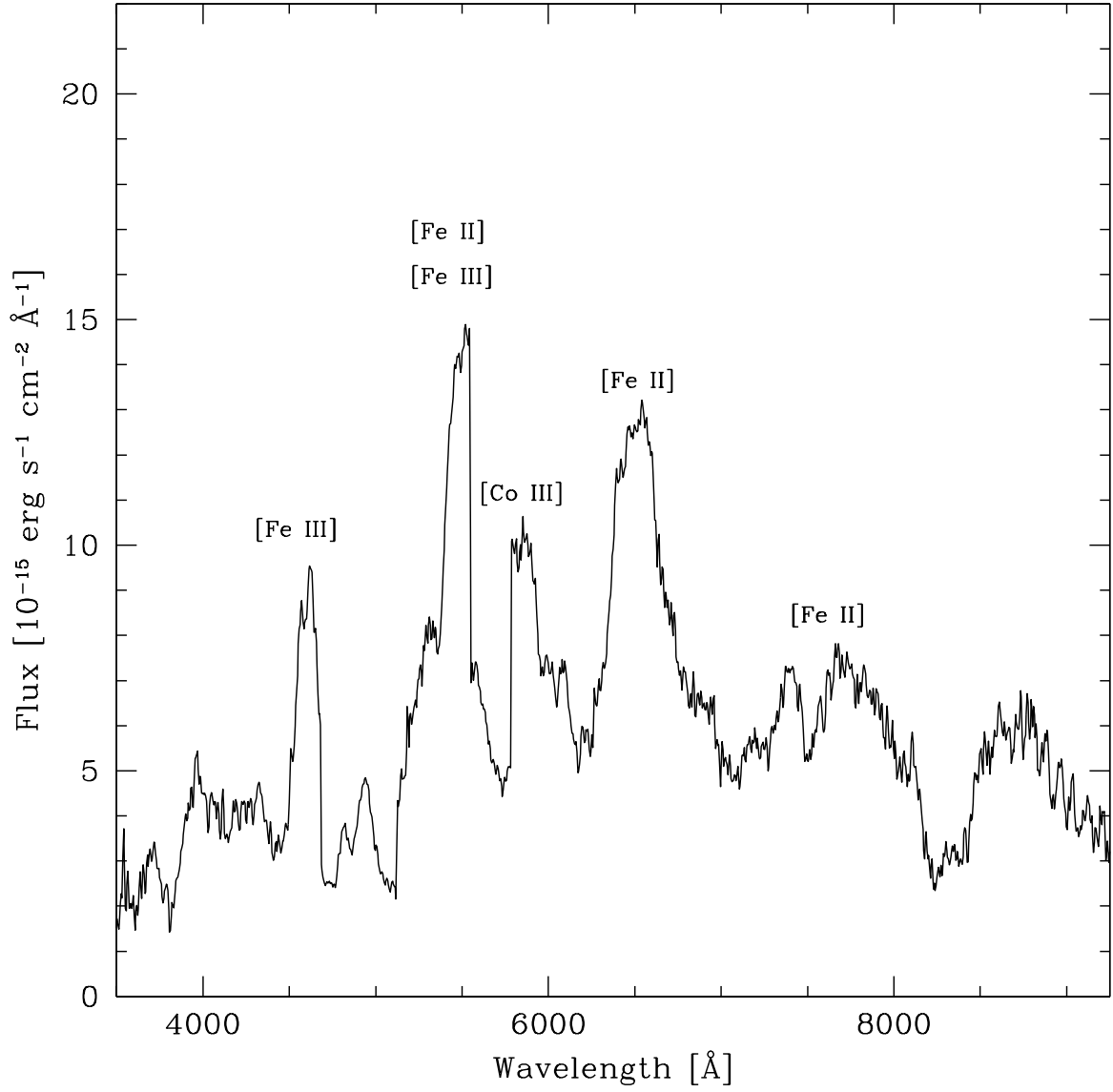}
\caption{Rest frame spectrum of the supernova obtained combining the 
data from the G430L and G750L gratings. 
\label{fig2}}
\end{figure}

\clearpage

\begin{figure}
\plotone{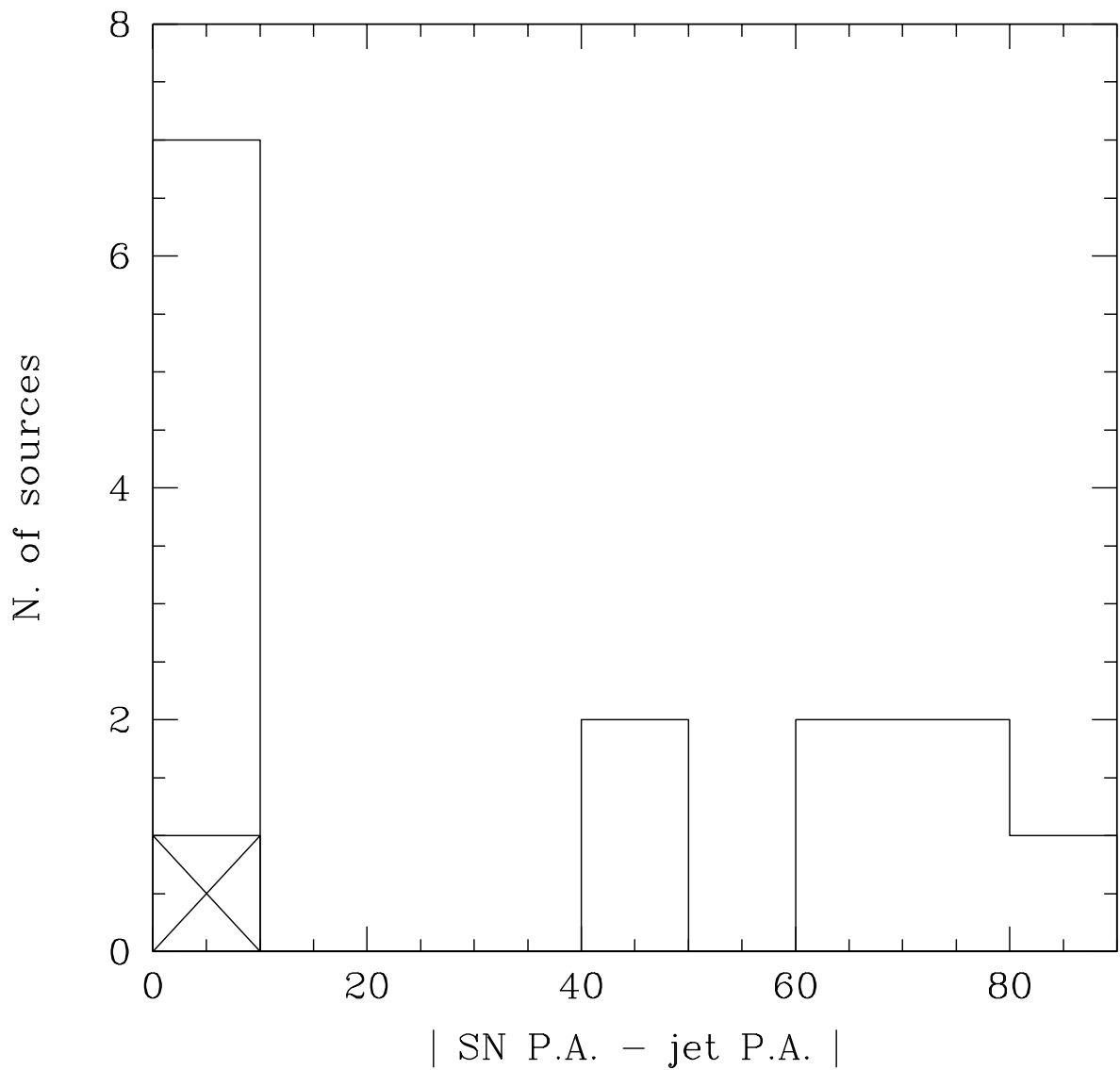}
\caption{Distribution of angular offset between 
the position angle of the 14 supernovae in radio-galaxies with respect to the 
nearest radio-jet.
This histogram reveals a strong concentration of SN events close to the 
radio axis. The crossed symbol corresponds to SN 1981G in NGC 5127: 
although aligned with jet position angle it is not  
associated with it
being located at a distance of 14\arcsec \ from the nucleus, 
while the radio jet length in this source is $\sim$ 10\arcsec.}
\label{offset}
\end{figure}

\clearpage

\begin{deluxetable}{lllcrrrr}
\tabletypesize{\scriptsize}
\tablecaption{Historical supernovae in radio-galaxies \label{tbl-1}}
\tablewidth{0pt}
\tablehead{
\colhead{Host galaxy} & \colhead{Radio source}   & \colhead{SN name}   &
\colhead{SN Type} & \colhead{Offset [$\arcsec$]}
& \colhead{SN P.A. (deg.)}
& \colhead{Jet P.A. (deg.)}
& \colhead{SN-jet offset (deg.)} 
                            }
\startdata
NGC1218       & 3C 78        & SN 2000fs & Ia & 0.36W 0.40N & 42 & 51 & 9 \\
NGC1275       & 3C 84        & SN 1968A  & I  &   7E 24S   & 164  & 160 & 4  \\
NGC4261       & 3C 270       & SN 2001A  & Ia & 3W 11N     & -15  & 85  & 80 \\
NGC4374 (M84) & 3C 272.1     & SN 1957B  & Ia &   8W 47N   & -10  & 0   & 10 \\
              &              & SN 1980I  & Ia &  454E 20N  &  87  & 0  & 87  \\
              &              & SN 1991bg & Ia &   2W 57S   & -178 & -186& 8  \\
NGC4486 (M87) & 3C 274       & SN 1919A  & I  &  15W 100N  &  -9  & -70 & 61 \\
NGC4874       & B2 1257+28   & SN 1968B  & -- &  12W 1N    & -85  & -130& 45 \\
              &              & SN 1981G  & I  &  15E 10N   & 56   & 50  & 6  \\
NGC5090     & PKS 1318-434 & SN 1981C  & -- &  15E 20S   & 143  & -138  & 79 \\
NGC5127       & B2 1321+31   & SN 1991bi & Ia &  13W 6S    & -115 & -65 & 50 \\
NGC5128 (CenA) & PKS 1322-42 & SN 1986G  & Ia & 120W 60S   & -117 & -121  & 4 \\
NGC5490       & PKS 1407+17  & SN 1997cn & Ia &   7E 12S   & 150  & 85  & 65 \\
NGC7503       & PKS 2308+07  & SN 2001ic & Ia &  15E  6N   & 68   & 66  & 2  \\
\enddata
\end{deluxetable}

\end{document}